\documentclass[aps,prb.reprint,twocolumn]{revtex4} 
\usepackage{epsfig}
\usepackage{graphicx}
\usepackage{amsmath}
\usepackage{amsfonts}
\usepackage{amssymb}
\usepackage{ulem}
\usepackage{cancel}
\usepackage{graphicx}
 \usepackage{wrapfig}
\usepackage{epsfig}
\usepackage{graphicx}
\usepackage{amsmath}
\usepackage{amsfonts}
\usepackage{amssymb}
\usepackage{graphicx}
 \usepackage{wrapfig}
\usepackage{textcomp}
\usepackage{pgf,pgfarrows,pgfnodes,pgfautomata,pgfheaps,pgfshade}
\usepackage{amsmath,amssymb}
\usepackage[colorlinks=true, linkcolor=blue]{hyperref}
\newcommand{{\sign}}{\rm sign}

\begin{document}

\title{
Excess equilibrium noise in topological SNS junction between chiral Majorana liquids  
}
\author{Dmitriy S. Shapiro$^{1,2,3 }$} \email{shapiro.dima@gmail.com} 
\author{Alexander D. Mirlin$^{4,5,6}$} \author{Alexander Shnirman$^{4,5}$}
\affiliation{$^1$Dukhov Research Institute of Automatics (VNIIA),  Moscow 127055, Russia}
\affiliation{$^2$L. D. Landau Institute for Theoretical Physics, Russian Academy of Sciences, Moscow 117940, Russia}
\affiliation{$^3$V. A. Kotel'nikov Institute of Radio Engineering and Electronics, Russian Academy of Sciences, Moscow 125009, Russia}
\affiliation{$^4$Institut f\"ur Nanotechnologie, Karlsruhe Institute of Technology, 76021 Karlsruhe, Germany}
\affiliation{$^5$Institut f\"ur Theorie der Kondensierten Materie, Karlsruhe Institute of Technology, 76128 Karlsruhe, Germany}
\affiliation{$^6$Petersburg Nuclear Physics Institute,  St.Petersburg 188300, Russia}

\begin{abstract}

We consider a Josephson contact mediated by 1D chiral modes on a surface of a 3D topological insulator with superimposed superconducting and magnetic layers. The system represents  an interferometer in which 1D chiral Majorana modes on the boundaries of superconducting electrodes are linked by ballistic chiral Dirac channels. We investigate the noise of the Josephson current as a function of the dc phase bias and the Aharonov-Bohm 
flux. Starting from the scattering formalism, a Majorana representation of the Keldysh generating action for cumulants of  the transmitted charge is found. At temperatures higher than the Thouless energy $E_{\rm Th}$, we obtain the usual 
Johnson-Nyquist noise, $4G_0 k_{\rm B}T$, characteristic for a single-channel wire with $G_0 \equiv e^2/(2\pi \hbar)$. At lower temperatures the behavior is much richer. In particular, the equilibrium noise is strongly enhanced to a temperature-independent  value $\sim G_0 E_{\rm Th}$ if  the  Aharonov-Bohm and superconducting phases are both close to $2\pi n$, which are points of emergent  degeneracy in the ground state of the junction. The equilibrium noise is related 
to the Josephson junction's impedance via the fluctuation-dissipation theorem.  In a striking contrast to usual Josephson junctions (tunnel junctions between two s-wave superconductors), the real part of the impedance does not vanish, reflecting the gapless character of Majorana modes in the leads.

\end{abstract}
\maketitle

\section{Introduction}

Noise of current as well as higher cumulants of charge fluctuations provide full information about quantum transport in  mesoscopic systems \cite{kogan2008electronic,nazarov2009quantum,BLANTER20001}. 
The theoretical technique of choice for the investigation 
of noise is the method of full counting statistics (FCS), introduced by Levitov and Lesovik \cite{levitov1993charge,levitov1996electron}
and adjusted later for the Keldysh description of transport in quantum circuits~\cite{nazarov2007full, PhysRevLett.88.196801}. 
This method allows for calculating the cumulant generating function (CGF) by means of a path integral with the generating term in the effective action. 
 It has been found that  the statistics of charge transfer is sensitive to electronic interactions which make individual tunnelling events correlated. Thus, the correlations may be probed by measuring the zero-frequency noise, i.e, the  transferred charge fluctuations normalised by a counting period.
For example, in a superconductor-normal metal-superconductor (SNS)  junction, the Cooper correlations in the terminals result,  in a certain range of parameters,  in a giant equilibrium noise \cite{PhysRevLett.76.3814,PhysRevB.53.R8891}. 
 This noise results from dissipative processes:  since the supercurrent flows in the ground state,  an ideal  Josephson junction  is noiseless. 

 During the  last decade, a considerable interest was generated by transport in topological  superconductors hosting    neutral  Majorana edge modes  \cite{alicea2012}.
The 1D Majorana chiral channels can appear in artificial hybrid structures based on 3D topological insulators (3DTI). As shown by Fu and Kane \cite{PhysRevLett.102.216403}, if one half of a surface of a 3D topological insulator is covered by an $s$-wave superconductor and  another half by a magnetic insulator, a gapless and chiral Majorana mode emerges at the border between the two coverings. Signatures of 1D gapless  Majorana modes were observed in STM spectroscopy of the Pb/Co/Si(111) structure~\cite{menard2017two}.

Combining magnetic and superconducting interfaces on top of 3DTI allows implementing new quantum interferometers. 
The  transport between normal metal terminals linked by the coherently propagating Majorana edge modes was studied in several papers. The     Mach-Zehnder devices    where Y-junction splits an electron   into   two Majorana fermions were  addressed in Refs. \cite{PhysRevLett.102.216403,PhysRevLett.102.216404}. The scattering theories of  Fabry-P\' erot and FCS of Hanbury Brown-Twiss interferometers were proposed in Ref.\cite{PhysRevB.85.125440} and Refs.\cite{PhysRevLett.107.136403,STRUBI2015489} respectively. It was shown that 3DTI based junctions with chiral  Majorana channels reveal an unusual interferometry and cross-correlation of noise in terminals.  Other realizations of chiral junctions with   Majorana egde modes  were proposed  in Refs.\cite{PhysRevB.83.100512,PhysRevB.83.220510,PhysRevB.93.161401,PhysRevB.88.075304}.  

In this work we study noise of the dc supercurrent carried through a single-channel link (normal part of the interferometer) which connects two 2D superconductors induced on the surface of a 3D topological insulator. The topological SNS junction under consideration is a quantum interferometer with chiral 1D Majorana liquids in the leads. The gapless nature of the leads provides an additional scattering channel along with the Andreev one. The Andreev states in the 1D normal wire can be viewed as scattering states     \cite{BLANTER20001}  of the incident Majorana fermions. This spectrum can be tuned into the degeneracy points by means of the gauge-invariant superconducting phase difference between the superconductors and by the Aharonov-Bohm phase in the normal channel. We show that this degeneracy leads to a strong enhancement of noise.

\begin{figure*}[ht] 
	\includegraphics[width=0.98\linewidth]{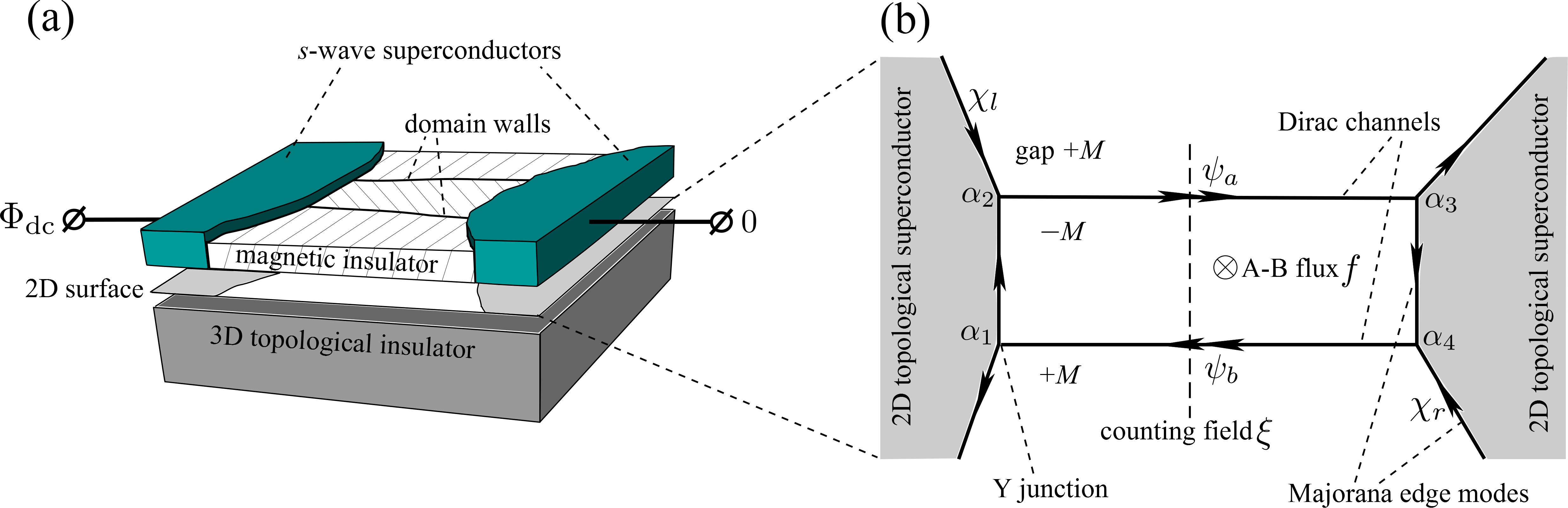} 
	\caption{(a) The chiral Josephson junction as a hybrid structure on a surface of a 3D topological insulator. 
		Two $s$-wave superconducting terminals, biased by the DC phase difference $\Phi_{\rm dc}$, have a proximity effect with the 2D Dirac surface.  The magnetic insulator film with two domain walls fills the space between the leads. 
				(b) The schematic view of the chiral 1D liquids on the 2D Dirac surface.  The  induced topological superconductivity under the $s$-wave terminals  is marked by gray color. The chiral Majorana liquids, marked by single arrows, are the edge modes supported by the interface of the superconducting and magnetic films. Magnetic domain walls  separate Zeeman gaps  of opposite signs $\pm M$ and support spinless  Dirac channels  $\psi_{a,b}$ marked  by double arrows.  
The direction of an arrow stands for chirality which depends on the direction of the magnetization. These channels are coupled by four Y-junctions with different  scattering phases $\alpha_i$ ($i=1,2,3,4$). The interferometer contour is threaded by the Aharonov-Bohm (A-B) flux $f$ resulting in the phase $\phi_{\rm AB}=\pi f/\Phi_0$ where $\Phi_0=h/(2e)$ is the superconducting flux quantum.  
The  Majorana channels $\chi_l$ and $\chi_r$ are the equilibrium incident modes. Their quantum fields are used in the  effective action. The Dirac complex fermions $\psi_a$ and $\psi_b$ describe the chiral Dirac channels connecting the two superconductors.
The counting variable $\xi$, generating the FCS of the transmitted charge, is introduced in the middle of the Dirac channels.} 
\label{setup} 
\end{figure*}

The studied Josephson setup and its schematic presentation in terms of a chiral SNS junction on a  2D surface of a 3D topological insulator are shown in Figs.~\ref{setup} (a) and (b) respectively. The chiral Dirac modes (the N part of the SNS) separate the areas of Zeeman gaps with the opposite signs $\pm M$. This setup, previously studied in Ref. \cite{PhysRevB.93.155411}, is based on the ideas of the  above mentioned  interferometers  \cite{PhysRevLett.102.216403,PhysRevLett.102.216404,PhysRevB.85.125440,PhysRevLett.107.136403,STRUBI2015489}, but it is actually dual to those interferometers because in our case the normal Fermi-liquid contacts are replaced by superconductors and Majorana edges while the interference loop involves the normal channels.

The equilibrium Josephson transport, the thermoelectric effect and the heat conductance controlled by the Aharonov-Bohm flux, enclosed by the chiral loop (see Fig. \ref{setup} (b)), were explored in \cite{PhysRevB.93.155411, PhysRevB.95.195425} for this system. The current-phase relationships might have  spikes or infinite derivatives for when both Aharonov-Bohm and the bias phases approach $2\pi n$, the points where the spectrum of Andreev states is highly degenerate. Generally, the density of states of the junction is continuous due to the coupling with the gapless contacts. At the degeneracy points the spectral current is rearranged and consists of singular points corresponding to discrete Andreev levels.     
   
The central results of this paper are related to the quantum regime, $T\ll E_{\rm Th}$, where the Thouless energy $E_{\rm Th}$ is inversely proportional to the dwell time in the interferometer loop. We show that the zero-frequency noise in this limit (i)  reveals periodic pattern as a function of superconducting and Aharonov-Bohm phases and (ii) is much larger (at the degeneracy points) than the thermal noise.

From the technical point of view, we use the method of FCS in order to calculate the fluctuations of the charge transmitted in the Dirac wires, where the definition of the current operator is straightforward. Our consideration is based on the FCS theory of  SNS  junction \cite{PhysRevLett.87.197006}, which is formulated  in terms of the Keldysh-Green functions in contacts and uses the scattering approach. The generating term (counting field) is introduced in the normal  link, as shown in Fig. \ref{setup} (b).
Taking into account that the  Dirac fermions in the normal channels are enslaved to the scattering states of incident Majorana modes, the generating Keldysh action is formulated in terms of Majorana variables and their equilibrium Green functions. 
We derive a generalized Levitov-Lesovik formula for CGF and, after that, the zero-frequency noise is obtained by performing the second order expansion  in the  quantum component of the counting field.
   
    The paper is organized as follows. In Sections  \ref{Chiral_Josephson_junction} and \ref{Scattering_approach}
   the chiral Josephson junction  and  scattering formalism are introduced. In Section \ref{Generating_action} the generating Keldysh action for cumulants is presented. In Section \ref{Integration} the path integral is calculated and an expansion for the cumulant generating function is obtained. In Section \ref{Results} the results for zero-frequency noise and current are derived with the use of the FCS. Specifically, in Sec.~\ref{Results:General_expressions} the general expression for zero-frequency noise is presented, in Sec~\ref{Results:Equilibrium_current} the result of Ref.~\cite{PhysRevB.93.155411} for the equilibrium current
is rederived, while in Sections \ref{Results:Equilibrium_noise_Low_temperature}, \ref{Results:Equilibrium_noise_Arbitrary_temperature}, and \ref{Results:Equilibrium_noise_High_temperature} the  noise is calculated for the low, intermediate and high temperature regimes, respectively. In Section \ref{Discussion} we summarize and discuss obtained results.

\section{Chiral Josephson junction}
\label{Chiral_Josephson_junction}

We assume the temperature to be much lower than the superconducting ($\Delta$) and magnetic ($M$) gaps induced in the 2D surface, $T<\min\{\Delta, M \}$ (see Fig.~\ref{setup}).  This allows us considering the system as a circuit of 1D wires and 
using the 1D scattering formalism. The transport of Cooper pairs under the phase bias $\Phi_{\rm dc}$  is governed by interference between the fermions in the neutral (Majorana) and charged (normal, Dirac) 1D channels with energies in the subgap domain.  The  scattering matrix of Andreev processes is modified by the presence of the additional reflection channel into the Majorana edge mode (see Fig.~\ref{setup}). 
This is an important distinction from a conventional SNS contact: the charged modes in the normal link are  the scattering states, or superpositions, of the incident Majorana fermions. In particular, the gapless nature of incident  modes results in a considerable thermal conductance (see Ref.~\cite{PhysRevB.95.195425}), which is in contrast to its exponential suppression due to the quasiparticle gap in regular SNS contacts. Also, it is important that Andreev pairs are non-local in the split Dirac channels. Hence,  the magnetic flux $f$ threading the area between the Dirac channels induces the single-electron Aharonov-Bohm phase $\phi_{\rm AB}=2\pi e f/h $. 
The  period of the critical current pattern is doubled as compared to conventional SQUIDs, i.e. it is given by $2\Phi_0=h/e$. Besides, the geometric asymmetry of contacts in the split normal channel together with the broken time-reversal symmetry  allows for the ``$\varphi$-junction effect" (Ref.~\cite{PhysRevB.67.220504}). Namely, a non-zero supercurrent can flow without a phase bias or temperature gradient.

The incident chiral Majorana mode emerges in the 2D surface at an  interface between superconducting and Zeeman gaps \cite{PhysRevLett.102.216403}. The Hamiltonian under consideration involves the spin-orbital interaction term of Rashba or Dresselhaus type and two terms related to the proximity effect of magnet ($x>0$ half-plane) and $s$-wave superconductor ($x<0$ half-plane). The corresponding solution of the Bogolyubov-de Gennes equation was found to be non-degenerate. The latter means that the Bogolyubov  quasiparticle operator corresponding to this solution is a real Majorana fermion with \begin{equation}
	\hat\chi_k=\hat\chi_{-k}^+. \label{chi-chi-dagger}
\end{equation}
	 The dispersion of this mode reads $\varepsilon_k=[{\rm sign} M] v k$ where  the direction  depends on the  sign of the Zeeman field and $v$ is the Fermi velocity of the 2D Dirac surface. The   chirality and gapless dispersion follows  from the broken time reversal symmetry by the  magnet. In contrast to Bogolyubov quasiparticles in conventional superconductors, the chiral Majorana excitation of a momentum $k$ is composed of four components: electron-like states of $k$ and $-k$ with the opposite spin orientations  and their hole-like counterparts. The neutrality  follows from the fact that four components of the Nambu eigenfunction have equal magnitudes for all the subgap  energies.  
The gapless dispersion means that the Majorana quasiparticles are excited at arbitrary low temperatures in contrast to the case of a conventional superconductor with a quasiparticle gap. The isolated Majorana edge mode transfers energy with the heat conductance which equals to 1/2 of that of the normal single  channel because the particle and antiparticle are not independent excitations. The Majorana mode is electrically neutral  and does not carry charge  or possess charge fluctuations.  Nevertheless, the superposition of two Majorana fermions mixed in tunnel or Y-junctions transports charge  and, consequently, generates charge noise. 

In our setup (Fig. \ref{setup} (a)) there are two $s$-wave superconducting terminals, biased by the DC phase difference $\Phi_{\rm dc}$, which both cover the surface of 3DTI.  The space between the leads is filled with magnetic insulator film with two domain walls. The superconductor/magnetic insulator structures and domain walls allows to implement the interferometer  consisting of four Y-junctions in 2D surface, see Fig. \ref{setup} (b). 
The proximity  induced topological   superconductivity in the helical 2D Dirac states is marked by gray color with the Majorana edge modes are marked by single arrows. The direction of each arrow stands for chirality which depends on the magnetization sign. Magnets induce Zeeman gaps of different signs leading to the emergence of two chiral channels with spinless Dirac fermions  (marked by double arrows). They form the split single channel Josephson link. The Cooper pairs are carried through these two chiral Dirac channels via the non-local Andreev pairs. The typical widths of the guiding channels for Majorana and Dirac modes are  given by coherence lengths in gapped magnetic or superconducting sectors.
All the 1D modes are spinless, due to the spin-momentum locking in the Dirac cone, meaning that there is no spin degeneracy, and that the wave function in the guiding channels has a spin texture. The latter results in non-trivial effects of Berry phase on the scattering phases $\alpha_i$ in the Y-junctions' nodes. The single electron Aharonov-Bohm phase $\phi_{\rm AB}$ is induced by an external  magnetic flux $f$ threading the interferometer bar.

In order to formulate the scattering approach we introduce the Hamiltonians  of the Majorana and Dirac chiral liquids, $\hat H_M$ and $\hat H_D$. They describe coherent propagation of  neutral and  charged 1D fermions and are derived from the Gor'kov-Nambu Hamiltonian for the 2D Dirac surface in proximity with magnets and superconductors. For the chiral Majorana liquid one has
	\begin{equation} 
		\hat H_{M}=\frac{v}{2}\sum\limits_{k}k\hat \chi    ( -k) \hat \chi ( k)\ ,
	\end{equation}
where the real fermion operators obey the condition (\ref{chi-chi-dagger}).
The prefactor of $1/2$ is the consequence of the excitations with momenta $k$ and $-k$ being non-independent. 
For the chiral Dirac mode we have a standard chiral Hamiltonian with complex fermions $\psi(k) \neq \psi(-k)^+$
	\begin{equation} 
	\hat H_{D}=v\sum\limits_{k}k\hat \psi^+(k) \hat \psi ( k).
\end{equation}
The Bogolyubov operators $\hat \psi(k)$ and $\hat \chi(k)$ and the effective 1D Hamiltonians follow from the solution of the Bogolyubov-de Gennes equation (see Ref. \cite{PhysRevB.95.195425} for  details). They involve the electronic and hole operators of bare 2D states in the surface.

\section{Scattering approach} 
\label{Scattering_approach}

Here we describe the scattering approach to the investigation of the interferometer. The detailed derivation can be found in Ref. \cite{PhysRevB.95.195425}.  Let us illustrate this derivation for the left contact between the superconducting area and the split Dirac channels. In the Y-junction $\#$1 an electron and a hole from the Dirac channel convert into two outgoing Majorana fermions (Fig.~\ref{setup} b). Incoming Majorana fermions are scattered in the Y-junction $\#2$ (Fig.~\ref{setup} b).  
The general form of the scattering matrix of the Y-junction $\#n$ is given by~\cite{PhysRevLett.102.216404}
 \begin{equation}
 \quad S_{in,\alpha_n}=\begin{bmatrix}
 1/\sqrt{2} &&  1/\sqrt{2}\\ \\ {\rm i}/\sqrt{2}  && -{\rm i}/\sqrt{2}
 \end{bmatrix}\begin{bmatrix}
 e^{{\rm i}\alpha_1} &&  0\\ \\ 0  && e^{-{\rm i}\alpha_n}
 \end{bmatrix}.\label{s-in}
 \end{equation}
where phase $\alpha_n$ depends on the microscopic details of the junction. For the Y-junction $\#$2 we get $S_{out}=S_{in}^T\label{s-out}$. The value of $\alpha_n$ is left arbitrary and is assumed to be independent of the momentum $k$ of the scattered particles. 

We eliminate the Majorana leg of length $d$, which connects between the Y-junction nodes  $\#$1 and $\#$2 and in which the dynamic phase $kd$ is acquired. As a result we obtain the S-matrix of the left combined chiral contact. This S-matrix 
describes the scattering of one incoming Dirac and one incoming Majorana modes into one outgoing Dirac and one outgoing Majorana modes. It acts on the vector of three amplitudes. The vector consists of the incident electron, Majorana fermion, and hole fields, $(\hat\psi_{in, k}, \ \hat\chi_{in, k}, \ \hat\psi^+_{in, -k})^T$. The outgoing fields are thus obtained through the S-matrix as
   \begin{multline}
   \begin{bmatrix}
   \hat\psi_{l,out, k} \\ \\
   \hat\chi_{l,out, k} \\ \\
   \hat\psi^+_{l,out, -k} \\
   \end{bmatrix}
   = 
   \\
   =\begin{bmatrix}
   \frac{1}{2} e^{{\rm i} kd +{\rm i}( \text{$\alpha_1 $}+\text{$\alpha_2 $})} & \frac{{\rm i }e^{{\rm i }\text{$\alpha_2 $}}}{\sqrt{2}} & \frac{1}{2} e^{{\rm i }kd -i( \text{$\alpha_1 $}-\text{$\alpha_2 $})} \\ \\
   \frac{{\rm i }e^{{\rm i }\text{$\alpha_1 $}}}{\sqrt{2}} & 0 & -\frac{{\rm i }e^{-i\text{$\alpha_1 $} }}{\sqrt{2}} \\ \\
   \frac{1}{2} e^{{\rm i }kd  + i( \text{$\alpha_1 $}-\text{$\alpha_2 $})} & -\frac{{\rm i }e^{-{\rm i }\text{$\alpha_2 $} }}{\sqrt{2}} & \frac{1}{2} e^{{\rm i }kd  - i( \text{$\alpha_1 $}+\text{$\alpha_2 $})} \\
   \end{bmatrix}
   \begin{bmatrix}
   \hat\psi_{l,in, k} \\ \\
   \hat\chi_{l,in, k} \\ \\
   \hat\psi^+_{l,in, -k} \\
   \end{bmatrix}\ , \label{s-matrix}
   \end{multline}
   where the subscript $l$ refers to the left contact.
To account for the non-zero superconducting bias phase $\Phi_{\rm dc}$ of the left superconducting contact, we employ the transformation  $\hat\psi_l\to e^{i\Phi_{\rm dc}/2}\hat\psi_l$.

The left and right leads confine the Dirac channels and the electronic states in the normal region become 
enslaved to the incident Majorana modes  in the left and right leads. Due to the partial Andreev reflection  the electrons and holes of momenta $k$ and $-k$ are not independent.
After some algebra we find a linear non-unitary transformation $\mathbf{R}_k$, 
\begin{equation} \label{eq:Rmatrix}
\begin{bmatrix} 
		\hat\psi_a(k)\\  
		\hat\psi_b(k)
	\end{bmatrix}=\mathbf{R}_k \begin{bmatrix} 
		\hat\chi_l(k)\\  
		\hat\chi_r(k)
	\end{bmatrix} \ , 
\end{equation}
which relates the fermionic fields in the middle of the lower and upper Dirac channels $\hat\psi_a$ 
and $\hat\psi_b$ to the fields $\hat\chi_l$ and $\hat\chi_r$ of the incident channels, as  shown in the Fig.\ref{setup} (b).  
The    matrix $\mathbf{R}_k$ is derived by eliminating the outgoing Majorana modes from the left and right lead's S-matrix relations (\ref{s-matrix}). A $k$-dependent dynamical phases in the chiral Dirac links and the Aharonov-Bohm and superconducting phases are included in the matrix elements.
For further convenience we introduce the phases $\Phi$ and $\phi$  which include the  scattering phases $\alpha_i$. For the dc phase bias this is akin to a ``$\varphi$-shift" (Ref.~\cite{PhysRevB.67.220504,PhysRevB.96.165422}), i.e.,
\begin{equation}  
\Phi=\Phi_{\rm dc}+ \frac{1}{2}(\alpha_1-\alpha_2-\alpha_3+\alpha_4)\ .
\end{equation}
For the Aharonov-Bohm phase the shift reads
$$
\phi=\phi_{\rm AB}+(\alpha_1+\alpha_2+\alpha_3+\alpha_4)\ ,
$$
see \cite{PhysRevB.95.195425} for details.
For the rectangular setup of Fig. \ref{setup} (b) the matrix $\mathbf{R}_k$ is parametrized by the coefficients $r_1(k, \phi, \Phi)$ and $ r_2(k, \phi, \Phi)$ as follows
\begin{equation}\label{eq:Rk}
	\mathbf{R}_k= 
	\begin{pmatrix} 
		r_1(k, \phi, \Phi) && r_2(k, \phi, \Phi)\\ \\ 
		r_2(k, \phi,- \Phi) && r_1(k, \phi, -\Phi)
	\end{pmatrix}.
\end{equation}
We obtain
\begin{equation}
	r_1(k, \phi, \Phi)=\frac{i \sqrt{2}e^{\frac{1}{4} i (2 L k +\phi- \Phi  )} \left(1+e^{i (\Phi   +\phi )}-2 e^{i (\phi-\varphi_k)}\right) }{ 1+2e^{i   \phi }\cos \Phi +e^{2i   \phi}  -4 e^{i  (\phi-\varphi_k) }}   ,
\end{equation} 
\begin{equation}
	r_2(k, \phi, \Phi)=\frac{2\sqrt{2}e^{\frac{1}{4} i (\frac{2 L k}{v}-2\varphi_k +5\phi- \Phi )} \sin\left(\frac{\Phi+\phi}{2}\right) }{ 1+2e^{i   \phi }\cos\Phi+e^{2i   \phi}  -4 e^{i  (\phi-\varphi_k) }}   ,  \label{eq:r1r2}
\end{equation}
where the dynamic phase and the Thouless energy are given by
$$
\varphi_k=\frac{\hbar v k}{E_{\rm Th}}\ ,\  E_{\rm Th}=\frac{\hbar v}{2d+2L}.
$$
 
At a first glance, relation (\ref{eq:Rmatrix}) indicates the reduction of the number of degrees of freedom, making operators of scattered electrons and holes $\hat\psi_{a,b}^+(-k)$ and  $\hat\psi_{a,b} ( k)$ not independent of each other. This is, however, not the case because it rather reflects a {\textit { rearrangement}}  of the degrees of freedom with their number being conserved (see a discussion in Sec.~\ref{Section:EqNoiseLowTemp}). 
 
The variety of the interference patterns, encoded by $r_1$ and $r_2$, influence the transport of Cooper pairs. 
As shown previously \cite{PhysRevB.93.155411}, the interference amplitudes define the spectral current which is $2\pi E_{\rm Th}$-periodic function of the energy and $2\pi$-periodic function of the dc phase bias $\Phi$. The remaining free parameter, the Aharonov-Bohm phase $\phi$,  modifies the width and the spectral shape of the spectral current. Analogously, a periodic in phases function will appear in the energy integral providing the value of the zero-frequency noise $S(\Phi,\phi,T)$.

\section{Generating Action}
\label{Generating_action}

In this section we use relation (\ref{eq:Rmatrix}) in order to find the effective Keldysh action which leads to the generating 
function for the FCS of the transmitted charge. We use the Majorana representation, for which the path integral is formulated in terms of the real (Majorana)  Grassmann variables. The matrix of the Green functions of the Majorana modes is diagonal in the channel space and correspond to the equilibrium free modes of momentum $k$.  
We introduce the counting field $\xi$ in the center of the Dirac counter propagating channels. This is a natural choice because the electric current is defined straightforwardly in the 1D chiral channels rather than in the  sectors covered by superconductors. Namely, the current operator is the difference between the chiral currents which flow in the upper and the lower Dirac channels
\begin{equation}
	\hat I = (-e)v (\hat\psi^+_a\hat\psi^{\phantom +}_a-\hat\psi^+_b\hat\psi^{\phantom +}_b)\ .
\end{equation}
Here $e$ is electron charge and $v$ is the Fermi velocity of the surface Dirac states.
Note that our method is distinct from that of \cite{PhysRevLett.87.197006}, where the counting field was inserted in one of the superconductors and the generating term was gauged out from the action by means of a transformation of the Green function of the corresponding lead. 

The cumulants of the
transported charge $N$ during the counting time $0<t<t_0$ are given by  the logarithmic derivatives of the corresponding partition function $\mathcal{Z}[\xi]$.  Namely, the CGF is given by  
\begin{equation}
{\rm CGF}(\xi) =  \ln \frac{\mathcal{Z}[\xi]}{\mathcal{Z}[0]}. \label{cgf-0}
\end{equation}
with the cumulants are 
\begin{equation}
	C_n=(-{\rm i})^n\left. \frac{\partial^n {\rm CGF}(\xi)}{\partial \xi^n}\right|_{\xi=0}. \label{cumulants-def}
\end{equation}
 The partition function is given by the path integral with the time-ordered exponent \cite{Kamenev} along the Keldysh contour 
 $\mathcal{C}$,
\begin{equation}
	\mathcal{Z}[\xi ]=  \langle\mathcal{T} e^{{\rm i} \int_\mathcal{C} \frac{\sigma_z}{2}\xi(t) I(t){\rm d}t }\rangle . \label{z-0}
\end{equation}
The variable $\xi$ is the amplitude of counting field $\xi(t)$ which is fully quantum in terms of Keldysh formalism, i.e. we should take $\sigma_z=+1$ for the forward and $\sigma_z=-1$ for backward parts of the contour. 
Moreover 
\begin{equation}\xi(t )=  \xi \theta(t )\theta(t_0-t )\ ,\end{equation}
i.e., the counting field is switched on and off at $t=0$ and $t=t_0$ respectively. Upon transition to the 
physical time $t$, the quantum counting field $\xi(t)$ is coupled to the classical component of the current defined as
$$
I_{\rm cl}(t)=\frac{I(t_+)+I(t_-)}{2}\ ,
$$
where $t_+$ and $t_-$ represent the physical time $t$ at the upper and lower branches of the Keldysh contour respectively.

For the averaging in (\ref{z-0}) we need the fermionic action describing the dynamics of the 
Dirac fields $\psi_{a,b}, \bar\psi_{a,b}$ in the split normal channel. As mentioned above, 
it is most natural to express these fields in terms of the two incident Majorana variables $\chi_{l,r}$
and to perform the path integration in terms of these Grassmann variables.
After this transformation the current $I_{\rm cl}(t)$ becomes a non-diagonal object.

The diagonal action  $\mathcal{S}_0 $ for the incident Majorana fermions reads on the Keldysh contour
\begin{equation} 
	\mathcal{S}_0=\sum\limits_{k,\gamma=L/R}\int_\mathcal{C} {\rm d}t \ 
	\left(\frac{{\rm i}}{2} \chi_{\gamma}( -k)\partial_t\chi_{\gamma}( k) - H_{\gamma} \right) \label{s-xi-0}
\end{equation}
with the Hamiltonians
\begin{equation} 
	H_{\gamma}=\frac{v}{2}\sum\limits_{k}k\chi_{\gamma}   ( -k)  \chi_{\gamma}( k).
\end{equation}
The factor of $1/2$ in time derivative of $\mathcal{S}_0$ is because we are dealing with real fermions. Thus we obtain
\begin{equation} 
	\mathcal{S}_0=\frac{1}{2}\sum\limits_{k,\gamma=l,r }\int_\mathcal{C} {\rm d}t \ 
	\chi_{\gamma} (t, -k) G_{\gamma,\gamma}^{-1}(t,t',k) \chi_{\gamma}( t',k)\ , \label{s-xi-0}
\end{equation}
where $G^{-1}_{i,j}(t,t',k)$ is the equilibrium inverse Green's function of usual charged fermions.

For the partition (generation) function we obtain  
\begin{equation}
	\mathcal{Z}[\xi ] = \int{\rm D}[X ] \exp \left({\rm i} \mathcal{S}_\xi [X ]\right) , \label{z-1}
\end{equation}
where 
the corresponding Grassmann fields after the Keldysh rotation are collected in the vector $X_k(t)$
$$
X^T_k(t)=[\chi_{1,L}(t,k); \chi_{2,L}(t,k); \chi_{1,R}(t,k) ; \chi_{2,R}(t,k) ]\ .
$$
Here the first index ($1,2$) indicates the Keldysh component, whereas 
the channel index $\gamma=l,r$ stands for free modes incoming from the left/right leads.
The Keldysh rotation is defined as
\begin{eqnarray}
	\chi_1(t)=\frac{\chi_+(t)+\chi_-(t)}{\sqrt{2}} , \quad  \chi_2(t)=\frac{\chi_+(t)-\chi_-(t)}{\sqrt{2}}\ ,
	\label{def-1-2-chi}
\end{eqnarray}
where $+ / -$ stand for direct/inverse branches of  $\mathcal{C}$-contour.
The Gaussian integration over the real (Majorana) Grassmann variables gives 
$$\int {\rm D}[X]\exp\left(-\frac{1}{2}X^T \hat A \  X\right)=\sqrt{{\rm Det} \hat A }\ .$$
The action $\mathcal{S}_\xi$ in (\ref{z-1}) reads
\begin{multline} 
	\mathcal{S}_\xi[X ]=\\
	\sum\limits_{k,p} \!\int\!\! {\rm d} t {\rm d} t' X^T_{-p}(t)\!\!\left[ \frac{1}{2}\delta_{p,k}  \check{\mathbf{G}}_{k}^{-1}(t,t'){+}\xi (t)\delta(t{-}t') 
	  \check{\mathbf{J}}_{p,k}\right]\!\! X_k(t'). \label{s-xi}
\end{multline}
This is  a sum of  the action $\mathcal{S}_0[X ]$ of the incident  Majorana channels and the generating term $\xi(t) \check{\mathbf{J}}_{p,k}$.  The bold font stands for the  matrix structure in channel space and  "check"-  symbol means the Keldysh space.
The  structure of the  matrix $ \check{\mathbf{J}}_{p,k}$, which parametrizes the current in the center of normal channels via the fields $X_k$, is to be derived with the help of $\mathbf{R}_k$-matrix.  

At this step we define the  matrix Green function in  (\ref{s-xi})
\begin{equation}
	 \check{\mathbf{G}}_k(t,t')=
	\begin{bmatrix}
		 \check{G}_{l}(t,t',k) && 0 \\ \\ 
		0 &&  \check{G}_{r}(t,t',k)
	\end{bmatrix}. \label{g-loc-0}
\end{equation}
The diagonal blocks $\check G_{l,r}$ are the Keldysh Green's functions
\begin{equation}
	\check{G}_{l}(t,t',k)=
	\begin{bmatrix}
		G_{l}^{\rm K}(t,t',k) && G_{l}^{\rm R}(t,t',k) \\ \\ 
		G_{l}^{\rm A}(t,t',k) && 0
	\end{bmatrix}, \label{g-l-p}
\end{equation}
\begin{equation}
	\check{G}_{r}(t,t',k)=
	\begin{bmatrix}
		G_{r}^{\rm K}(t,t',k) && G_{r}^{\rm R}(t,t',k) \\ \\ 
		G_{r}^{\rm A}(t,t',k) && 0
	\end{bmatrix}.  \label{g-r-p}
\end{equation}
We remind that these Green functions describe free chiral fermions. 
Let us introduce a Fourier representation of $X_{-k}(t)$ and $X_{k}(t)$ in (\ref{s-xi}) by the following rule 
\begin{equation}
	X_{-k}(t )=\int \frac{{\rm d}\omega'}{2\pi} X_{-k}(\omega' ) e^{{\rm i} \omega' t}, 
\end{equation}
\begin{equation}
	X_k(t)=\int \frac{{\rm d}\omega}{2\pi}   X_k(\omega) e^{-{\rm i} \omega t}.
\end{equation}
Then, the inverse Green function  from the action (\ref{s-xi})  is transformed into 
\begin{multline}
	\check{G}_{l,r}^{-1}(\omega,\omega',k)=\\
	=2\pi \delta(\omega'- \omega) \begin{bmatrix}
		2{\rm i}o (1-2n_{l,r}(k)) && \omega-vk+{\rm i} o \\ \\ 
		\omega-vk-{\rm i} o && 0
	\end{bmatrix}.  \label{g-inv}
\end{multline}
In turn, the  frequency representations of retarded (R), advanced (A) and Keldysh (K) components  in (\ref{g-l-p}, \ref{g-r-p}) read
\begin{equation}
	G_{l,r}^{\rm R}(\omega, k)=\frac{1}{\omega-vk+{\rm i} o}, \quad G_{l,r}^{\rm A}(\omega, k)=\frac{1}{\omega-vk-{\rm i} o}\ ,
\end{equation}
\begin{equation}
	G_{l,r}^{\rm K}(\omega, k)=-2\pi {\rm i}\delta(\omega - vk)(1-2n_{l,r}(k))\ .
\end{equation}
The only constraint for the distribution function $n_{L/R}(k)$ follows from the fact that Majorana mode $\chi$ is real. It reads
\begin{equation}
	n_{L/R}(k)=1-n_{L/R}(-k). \label{constraint}
\end{equation}
Next we calculate the current matrix $ \check{\mathbf{J}}_{p,k}$ acting in the basis of $X_k$. 
The definition for the current in terms of usual Grassmann variables reads
\begin{equation}
	I=(-e)v(\rho_{ a}-\rho_{b}), \quad \rho_{ a}=\bar\psi_a \psi_a, \quad  \rho_{b}=\bar\psi_b \psi_b\ . \label{i-1}
\end{equation}
The  classical and quantum components of the charge densities read
\begin{equation}
	\rho_{\rm cl} = \frac{1}{2}(\rho_+ + \rho_-)\ , \quad \rho_{\rm q} = \frac{1}{2}(\rho_+ - \rho_-)\ .
\end{equation} 
For the fermion variables we introduce the Keldysh indices 1 and 2 exactly as for $\chi$:
\begin{eqnarray}
	\psi_{1,2}(t)=\frac{\psi_+(t)\pm\psi_-(t)}{\sqrt{2}} , \nonumber\\ 
	\bar\psi_{1,2}(t)=\frac{\bar\psi_+(t)\pm\bar\psi_-(t)}{\sqrt{2}}\ . 
	\label{def-1-2}
\end{eqnarray}
Using (\ref{def-1-2}) we obtain 
\begin{multline}
	I=
	\frac{(-e)v}{2}\left(\bar\psi_{a,1}\psi_{a,1}+\bar\psi_{a,2}\psi_{a,2}\right. \\ \left. -\bar\psi_{b,1}\psi_{b,1}-\bar\psi_{b,2}\psi_{b,2}\right)\ ,  \label{j-0}
\end{multline}
which can be rewritten as
\begin{equation}
	I(t)=\sum\limits_{p,k}\sum\limits_{\gamma,\sigma}\bar\psi_{p;\gamma,\sigma}(t)\check {\mathbf{ I }}_{\gamma,\sigma}\psi_{k;\gamma,\sigma}(t)\ .
\end{equation}
Here $\sigma=1,2$ is the Keldysh index and 
\begin{equation}
	\check {\mathbf{ I }}=-\frac{ev}{2}   \gamma^z \sigma^0\ ,
\end{equation}
where $\gamma^z$ is the Pauli matrix in the channel space ($a/b$) and $\sigma^0$ is the unity matrix in the Keldysh space.

At this step we introduce the complex Dirac field $\Psi_k$ in the extended Gor'kov-Nambu $\tau$-space 
\begin{multline}
	\Psi^T(t,k)=[\psi_{a,1}(t,k),\psi_{a,2}(t,k),\psi_{b,1}(t,k),\psi_{b,2}(t,k), \\
	\bar\psi_{a,1}(t,k),\bar\psi_{a,2}(t,k),\bar\psi_{b,1}(t,k),\bar\psi_{b,2}(t,k)]\ .
\end{multline}
Such an extension is necessary in order to take into account superconducting correlations of Dirac fermions.
In terms of these fields the current now reads 
\begin{equation}
	\check{\mathbf I}_\tau(t)=\sum\limits_{p,k} \bar\Psi^T (t,p)\frac{\tau^z}{2}\check{\mathbf{ I }}\Psi (t,k) .  
	\label{i-3}
\end{equation}
The relation between scattered Dirac and incoming Majorana modes is given by Eq.~(\ref{eq:Rmatrix}).
The extension of $\mathbf{R}_k$ to the Keldysh $\sigma$-space requires a simple direct product with $\sigma^0$. The relation between the 4-dimensional $X_k$ and the 8-dimensional $\Psi_k$ reads
\begin{equation}
	\Psi(t,k)=\begin{bmatrix}	
		\sigma^0\mathbf{R}_k X_k \\
		\sigma^0\mathbf{R}^*_{-k} X_{k}
	\end{bmatrix}\ . \label{psi-x-relation}
\end{equation}
The Hermitian conjugation of the (\ref{psi-x-relation}) reads
\begin{equation}
	\bar\Psi_k^T(t,k)=\begin{bmatrix}	
		X_{-k}^T \sigma^0\mathbf{R}_k^+  &&  X_{-k}^T\sigma^0\mathbf{R}^T_{-k} 
	\end{bmatrix}. \label{psi-x-relation-T}
\end{equation}
Using Eq. (\ref{i-3}) we obtain the expression for the current $I_{\rm cl}(t)$ written in the $X$ basis. The Nambu $\tau$-index is trivially traced out and the current in Majorana basis now reads
\begin{equation}
	I_{\rm cl}(t)=\sum\limits_{p,k}  X_{-p}^T (t)\check{\mathbf{J}}_{p,k}X_k(t)\ ,  
	\label{i-4}
\end{equation}
where the kernel matrix $\check{\mathbf{J}}_{p,k}$ is obtained as follows
\begin{equation}
	\check{\mathbf{J}}_{p,k}  =-\frac{ev}{4}(\mathbf{R}_p^+ \gamma^z \mathbf{R}_k-\mathbf{R}_{-p}^T \gamma^z \mathbf{R}_{-k}^*)\sigma^0.
	\label{i-4}
\end{equation}
This matrix provides the generating counting term in the Majorana representation.

\section{Integration over Majorana fields}
\label{Integration}
In this section we perform the integration over the Majorana field $X_k(t)$ in the path integral (\ref{z-1}). We start from a transformation of the action (\ref{s-xi}) into a frequency integral. After that it is transformed into a discrete sum with a step of $\Delta\omega $ and, finally, the CGF is found. 
The generating part of the action $\mathcal{S}_\xi$ (\ref{s-xi}) with $\xi(t)=\xi \theta(t)\theta(t_0 - t)$
transforms as follows in the frequency representation
\begin{multline}
	\int   t X^T_{-p}(t)\xi (t) 
	\check{\mathbf{J}}_{p,k} X_k(t) {\rm d}t \\= \xi \int  f(\omega' , \omega)  X^T_{-p}(\omega') 
	\check{\mathbf{J}}_{p,k} X_k(\omega) \frac{{\rm d}\omega{\rm d}\omega'}{(2\pi)^2}. \label{int-t0}
\end{multline}
In the   integral of (\ref{int-t0}) we introduced $ f(\omega' , \omega)$ which is the Fourier transformation of  $\xi(t)$  
\begin{equation}
	f(\omega' , \omega)=\int\limits_0^{t_0} {\rm d}t \, e^{{\rm i} (\omega' - \omega) t}={\rm i} \frac{1-e^{{\rm i} (\omega' - \omega) t_0}}{\omega' - \omega}\ . \label{f}
\end{equation}
Discretization assumes that ${\rm d}\omega$ and ${\rm d}\omega'$ are replaced by $\Delta\omega$ and the delta function transforms into Kronecker symbol as $\Delta\omega \delta(\omega'-\omega)\to \delta_{\omega',\omega}$ and $ f(\omega' , \omega)$ is considered as matrix.
The generating  action now reads as
\begin{multline}
	\mathcal{S}_\xi[X ]= \sum\limits_{p,k;\omega',\omega}    X_{-p}(\omega' )\left[ \delta_{\omega' , \omega}\delta_{p , k}\frac{1}{2}  \check{\mathbf{G}}^{-1}(\omega,k) \frac{\Delta\omega}{2\pi}\right.  \\ 
	+ \left. \xi  f(\omega', \omega)\check{\mathbf{J}}_{p,k}  \left(\frac{\Delta\omega}{2\pi}\right)^2 \right] X_k(\omega). \label{s-xi-discrete}
\end{multline}
Calculation of the path integral with the discretized action from (\ref{s-xi-discrete}) and the definition (\ref{cgf-0}) gives the generalized Levitov-Lesovik formula for CGF of the contact
\begin{multline}
{\rm CGF}(\xi)=\\
	=\frac{1}{2}{\rm Tr} \ln \left[ \delta_{\omega' , \omega}\delta_{p , k}\gamma^0\sigma^0+ 2\xi  \check{\mathbf{G}}(\omega',p)  f(\omega', \omega) \check{\mathbf{J}}_{p,k}\frac{\Delta}{2\pi}  \right] . \label{lnz-2}
\end{multline}
The factor of 1/2 in (\ref{lnz-2}) results from square root of the determinant in integration over  real Grassman variables. 
The sign $\rm Tr$   assumes the trace taken over $p, k,\omega' ,\omega$ and $\sigma,\gamma$ indices. 
Calculation of the trace
  in a compact form is challenging  due to the non-diagonal structure in   momentum space  of the generating term $\check{\mathbf{J}}$ in the new Majorana basis. 
The formula  (\ref{lnz-2}) allows us to obtain the cumulants $C_n$  through the logarithm expansion up to the $n$-th order in $\xi$ and using the definition (\ref{cumulants-def}).  The second cumulant provides the central result of this paper for zero-frequency noise and is discussed in the next Section.

\section{ Results for zero frequency noise }
\label{Results}
\subsection{General expressions for the average current and the noise}
\label{Results:General_expressions}
In this section we obtain the general expression for the zero frequency noise of the current, $S$, which is the quantity of central interest of this work. The spectral density of noise $S_\omega(\phi,\Phi,T)$ is related to the symmetrized correlator as
\begin{equation}
	S_\omega(\phi,\Phi,T)=\int {\rm d}t \left(\langle I(t) I(0)\rangle+\langle I(0) I(t)\rangle- 2I^2\right) e^{{\rm i}\omega t}. \label{corr-def}
\end{equation}
The zero frequency value $S\equiv S_{\omega=0}$ is given by the second cumulant introduced above as 
\begin{equation}
	S=\lim\limits_{t_0\to \infty} 2\frac{C_2}{t_0}\ . \label{P0}
\end{equation}
In order to calculate $C_2$ we expand ${\rm CGF}(\xi)$ up to the second order in $\xi$ and transform  sums 
into integrals over frequencies:
\begin{multline}
{\rm CGF}(\xi)=\xi \sum\limits_k  \int \frac{{\rm d}\omega}{2\pi} f(\omega, \omega)  {\rm tr} \left[\check{\mathbf{G}}(\omega,k)    \check{\mathbf{J}}_{k,k} \right]  \\
	-\xi^2 \sum\limits_{k,p}\int   |f(\omega, \omega') |^2  {\rm tr} \left[\check{\mathbf{G}} (\omega,p )   \check{\mathbf{J}}_{p,k}\hat G (\omega',k )    J_{k,p} \right]   \frac{{\rm d}\omega{\rm d}\omega'}{(2\pi)^2} \ . \\ \label{lnz-3}
\end{multline}
Here $\rm tr$ denotes the trace over $\sigma$ and $\gamma$ indices only. In the integrand of the second term we took into account that $f(\omega, \omega')f(\omega', \omega)=|f(\omega, \omega') |^2$. 
For the first cumulant we obtain
\begin{multline}\label{eq:C1}
	C_1=t_0\frac{ev}{2}\sum\limits_k\left[(n_l(k)+n_r(k)-1) {\rm tr}_\gamma[ {\mathbf{R}}_k^+\gamma^z \mathbf{R}_k] \right.\\
	+\left. (n_l(k)-n_r(k)){\rm tr}_\gamma[\gamma^z \mathbf{R}_k^+\gamma^z \mathbf{R}_k]\right]\ ,
\end{multline}
where ${\rm tr}_\gamma$ denotes the trace over the channel indices only. 
The first term in (\ref{eq:C1}) is responsible for the Josephson current while the second one gives the thermoelectric effect discussed in \cite{PhysRevB.95.195425}. The corresponding  dimensionless spectral currents are given by 
${\rm tr}_\gamma[ {\mathbf{R}}_k^+\gamma^z \mathbf{R}_k]$ and ${\rm tr}_\gamma[\gamma^z \mathbf{R}_k^+\gamma^z \mathbf{R}_k]$ respectively.
Note that Eq.~(\ref{eq:C1}) was simplified by accounting for the constraint (\ref{constraint}) on the distribution functions of the Majorana fermions and using the limit of (\ref{f}), which gives 
$$
\lim_{\omega'\rightarrow\omega} f(\omega', \omega)=t_0\ .
$$

For the second cumulant we have $|f(\omega, \omega')|^2$ in the integrand. 
Assuming the measurement time $t_0$ is long, i.e., $t_0 \gg E_{\rm Th}^{-1}, T^{-1}$, we get the delta-function, i.e.,
$$
|f(\omega, \omega')|^2=\frac{2(1-\cos(\omega' - \omega)t_0)}{(\omega' - \omega)^2}\approx 2\pi t_0 \delta(\omega' - \omega)\ .
$$
Performing the integration over $\omega$ in the second line of (\ref{lnz-3}) and summation over $p$ in the continuous limit via    $\sum\limits_{p}\to \int\frac{{\rm d} p}{2\pi}$ we obtain 
\begin{multline}
	C_2=2t_0\sum\limits_{k } \left(\frac{4}{v}\right)  \left[\left( \check{\mathbf{J}}^{(1,1)}_{k,k}\right)^2n_l(k)n_l(-k)\right.  \\ +\left( \check{\mathbf{J}}^{(2,2)}_{k,k}\right)^2 n_r(k)n_r(-k)
	\\ \left. + \check{\mathbf{J}}^{(1,2)}_{k,k} \check{\mathbf{J}}^{(2,1)}_{k,k}[n_l(k)n_r(-k)+n_r(k)n_l(-k)]\right] , \label{c2}
\end{multline}
where the distribution functions $n_l(k)$ and $n_r(k)$ are arbitrary. 
The upper indices of $J_{k,k}$ are related to the channel space $\gamma$. 

For the particular case of identical distribution functions, $n(k)=n_l(k)=n_r(k)$, we obtain 
\begin{multline}
	C_2=t_0 e^2 v \sum\limits_{k }n(k)n(-k)  {\rm tr}_\gamma \left[\left(  \mathbf{R}_k^{+}\gamma_z \mathbf{R}_k \right)^2\right.  \\ - \left.   \mathbf{R}_k^{+}\gamma_z \mathbf{R}_k\left(  \mathbf{R}_{-k}^{+}\gamma_z \mathbf{R}_{-k} \right)^T\right] . \label{c2-equil}
\end{multline}

We have also performed an alternative derivation and obtained the same result for $C_2$ from a direct calculation of the noise $S_\omega$ using the operator approach. In this method we employ the relation   (\ref{eq:Rmatrix}) between the   Heisenberg operators $\hat\psi$ and $\hat\chi$ and insert them as linear combinations into  the definition for noise correlator $S_\omega$   (\ref{corr-def}).  In the course of the calculation of  $C_2$ we need the thermodynamic  averaging of the cumulants of four Majorana operators, which have the following form
\begin{multline}
	\langle\langle\hat\chi_{\gamma_1}(k_1)\hat\chi_{\gamma_2}(k_2)\hat\chi_{\gamma_3}(k_3)\hat\chi_{\gamma_4}(k_4)\rangle\rangle = \\
	\langle\hat\chi_{\gamma_1}(k_1)\hat\chi_{\gamma_4}(k_4)\rangle\langle\hat\chi_{\gamma_2}(k_2)\hat\chi_{\gamma_3}(k_3)\rangle \\- \langle\hat\chi_{\gamma_1}(k_1)\hat\chi_{\gamma_3}(k_3)\rangle\langle\hat\chi_{\gamma_2}(k_2)\hat\chi_{\gamma_4}(k_4)\rangle\ , \label{averaging}
\end{multline}
where $\gamma_i=l,r$ for $i=1,2,3,4$.
Comparing these two approaches we observe that all the terms given by the first trace  in (\ref{c2-equil})  are identical to the second line in  (\ref{averaging}). The last term of (\ref{c2-equil})  with the minus prefactor is given by the third line in  (\ref{averaging}).

\subsection{Equilibrium current}
\label{Results:Equilibrium_current}
In this subsection we calculate the cumulants $C_{1,2}$ in equilibrium, with the temperatures of the leads being equal to each other, $T_l=T_r=T$, so that the distribution function of the incident Majorana particles reads
\begin{equation}
	n(k)=\frac{1}{1+e^{\frac{vk}{T}}}\ .
\end{equation}
Replacing the sum over $k$ by the integral over the energy, $ \sum\limits_k \rightarrow \int \frac{{\rm d} \varepsilon}{2\pi \hbar v}$, we obtain
\begin{multline}
	C_1=t_0\frac{e}{4\pi\hbar}\sin\Phi\\ \times  \int  \frac{\sin\varphi_\varepsilon \tanh\frac{\varepsilon}{2T} \ {\rm d}\varepsilon }{1+\left(\frac{\cos \Phi+\cos \phi }{2}\right)^2- (\cos \Phi+\cos \phi )\cos\varphi_\varepsilon}\ , \\
\end{multline}
where the dynamical phase is labeled by the index $\varepsilon$:
$$
\varphi_\varepsilon=\frac{\varepsilon}{E_{\rm Th}} .
$$
Via the relation $I(\Phi,\phi)=C_1/t_0$ we arrive at
the current-phase relationship obtained previously in Ref. \cite{PhysRevB.93.155411}:
\begin{multline}
	I(\Phi,\phi) =  4\pi  {\frac{e k_{\rm B} T}{h} }\sin\Phi \\ \times   \sum \limits_{n=0}^\infty \frac{1}{2\exp\left(\pi \frac{k_{\rm B}T(1+2n)}{E_{Th}}\right)-\cos \phi-\cos\Phi }\ . \label{CPhR}
\end{multline}
As shown in Ref. \cite{PhysRevB.93.155411}, in the low-temperature limit, $k_{\rm B}T \ll E_{\rm Th}$, the summation can be replaced by the integration and the Josephson current shows non-sinusoidal oscillations as the function of the phases $\phi$ and $\Phi$ with the amplitude proportional to $E_{\rm Th}$:
\begin{equation}
	I(\Phi,\phi)_{T\ll E_{\rm Th}}= -\frac{e }{\pi \hbar }E_{\rm Th}  
				\sin(\Phi) \frac{\ln \left[1-\frac{\cos \phi +\cos \Phi}{2}\right]}{\cos \phi +\cos  \Phi}\ . \label{eq:I}
\end{equation}
This result exhibits an interesting singular behaviour near the points $\phi=2\pi n$, 
$\Phi=2\pi m$. In what follows we conclude that at these points the system shows large excess noise.

\subsection{Equilibrium noise:  General expression  
}
\label{Results:Equilibrium_noise_Low_temperature}

Using Eqs. (\ref{c2-equil}) and (\ref{P0}) we obtain the following expression for the equilibrium zero-frequency noise 
\begin{equation}
	S(\Phi,\phi,T)=G_0 \int  \frac{{\rm d}\varepsilon}{\cosh  \frac{\varepsilon }{T} +1} \mathcal{Y}(\Phi,\phi,\varphi_\varepsilon) \ ,\label{noise-0}
\end{equation}
where $G_0$ is the conductance quantum, $G_0=\frac{e^2}{2\pi\hbar}$. The kernel function is given by
\begin{multline}
	  \mathcal{Y}(\Phi,\phi,\varphi_\varepsilon) =\\=-4 \frac{A_0(\Phi,\phi)+A_1(\Phi,\phi)\cos\varphi_\varepsilon +A_2(\Phi,\phi)\cos 2\varphi_\varepsilon}{(B_0(\Phi,\phi)+B_1(\Phi,\phi)\cos\varphi_\varepsilon)^2} \label{y}
	  \end{multline}
with
\begin{multline}
	  A_0(\Phi,\phi)= (14 \cos \phi +\cos 3 \phi)  \cos \Phi\\+2  (\cos 2 \phi +6)\cos  2 \Phi +\cos 3 \Phi  \cos \phi -4 \cos 2 \phi -26; \\
	   A_1(\Phi,\phi)= -2  (\cos 2 \Phi +3 \cos 2 \phi +4)\cos \Phi \\+5 (5-2 \cos 2 \Phi ) \cos \phi +\cos 3 \phi;\\
	   A_2(\Phi,\phi)= 4(2-\cos 2 \Phi -\cos  2 \phi );\\
	   B_0(\Phi,\phi)= \cos 2 \Phi +4 \cos \Phi  \cos \phi +\cos 2 \phi +10;\\
	   B_1(\Phi,\phi)=-8(\cos \Phi +\cos \phi ).
\end{multline}
Despite the function $\mathcal{Y}(\Phi,\phi,\varphi_\varepsilon)$ being somewhat cumbersome, it can be simplified to a compact expression in the low, $T \ll  E_{\rm Th}$, and high, $T \gg  E_{\rm Th}$, temperature limits. Also, a certain simplification is possible for the degeneracy points $\Phi=2\pi m$, $\phi=2\pi n$, where an analytical calculation of (\ref{noise-0}) becomes possible for arbitrary temperatures. These three limits are discussed below.

It is important to note that the function $ \mathcal{Y}$ is essentially the spectral weight of fluctuations. 
Its integral over the period of the dynamical phase, $-\pi <\varphi_\varepsilon< \pi$, is independent of $\Phi$ and $\phi$: \begin{equation}
	\int\limits_{-\pi}^\pi \mathcal{Y}(\Phi,\phi,\varphi_\varepsilon) {\rm d}\varphi_\varepsilon=4\pi. \label{y0}
\end{equation}
In the limit of zero phases $\Phi=\phi=0$ (more precisely $\Phi=2\pi m$, $\phi=2\pi p$) the kernel is a sum of delta functions with singularities at $\varphi_\varepsilon=2\pi n$.  This is the limit of the degenerate ground state where the otherwise
continuous spectral weight  turns into a sum of discrete contributions due to the Andreev states at the energies $\varepsilon_{\rm A}=2\pi n E_{\rm Th}$. The invariance condition (\ref{y0}) provides the coefficient in front of the delta functions: 
\begin{equation}
	\mathcal{Y}(\Phi=\phi=0, \varepsilon)=  
		 4\pi E_{\rm Th}\sum\limits_{n=-\infty}^\infty \delta(\varepsilon-2\pi n E_{\rm Th})\ . \label{y-1}
\end{equation}

\subsection{   Equilibrium noise: Low temperature limit} 
\label{Section:EqNoiseLowTemp}

Our central results  follow from Eq.  (\ref{noise-0}) in the low temperature limit, $T\ll E_{\rm Th}$. The first result is the 
presence of the $\phi,\Phi$-dependent oscillations of the noise. The second one is the excess noise and, as a consequence, the large real part of the impedance of the system close to the degeneracy (as a representative point we take  $\Phi=\phi=0$). Namely, if the distance from the degeneracy point on the $\phi,\Phi$-plane is large, $ \sqrt{\phi^2+\Phi^2} \gg \sqrt{T/E_{\rm Th}}$, then we can expand the function $\mathcal{Y}$ given in (\ref{y}) around $\varphi_\varepsilon=0$ and we obtain
\begin{multline}
	S_{T\ll E_{\rm Th}}(\Phi,\phi) = 8 G_0T  \frac{(1-\cos\phi)(1+\cos\Phi)}{(2-\cos\phi-\cos\Phi )^2}   \\
	 +32\pi^2 G_0 T^3\frac{(\cos\phi-\cos\Phi)^3+8(2\sin^2\Phi-\sin^2\phi\cos\Phi)}{3E_{\rm Th }^2 (2-\cos\phi-\cos\Phi )^4} \\
	 +	 T G_0 O[ T^4/E_{\rm Th}^4], \quad \phi^2+\Phi^2\gg   \frac{T}{E_{\rm Th}} . \label{S-phi-low-temp}
	\end{multline}
The oscillations in (\ref{S-phi-low-temp}) show the usual $2\pi$-periodic pattern as a function of the superconducting phase 
$\Phi$ as expected for a system where the charge parity is not conserved. The dependence on the Aharonov-Bohm     phase $\phi$  is 
also $2\pi$ periodic, which corresponds to an unconventional for superconducting 
systems $h/e$ period in terms of the Aharonov-Bohm flux. This is due to the split chiral channels in our system and has been 
discussed in detail in Ref.~\cite{PhysRevB.93.155411}.
	
The leading term in  (\ref{S-phi-low-temp}) can be obtained from (\ref{noise-0}) by replacing 
the thermal distribution the delta function  as
\begin{equation}
	\frac{1}{\cosh  \frac{\varepsilon }{T} +1}\approx  2T\delta(\varepsilon).
\end{equation} 
This delta-functional approximation is valid, however, only far enough from the singularity point, i.e. if  
\begin{equation}\phi^2+\Phi^2\gg   \frac{T}{E_{\rm Th}}\ . \label{cr}
\end{equation}
In this case the distribution function constitutes a sharp peak of width $T$ compared to the smooth dependence of $\mathcal{Y}$ on the energy. Indeed, the function $\mathcal{Y}$ is peaked around $\varepsilon=0$ and 
the width of the  peak is given by
\begin{equation} 
	\Delta\varepsilon= E_{\rm Th} \frac{\phi^2+\Phi^2}{4\sqrt2}.  \label{w1}
\end{equation}
This follows from the expansion of the denominator of $\mathcal{Y}( \Phi,\phi ,\varphi_\varepsilon)$, which, 
if $\sqrt{\phi^2+\Phi^2} \ll \pi$, reads
$$
B_0(\Phi,\phi)+B_1(\Phi,\phi)\cos\varphi_\varepsilon\approx O\left[ \varphi_\varepsilon^2+\left(\frac{\phi^2+\Phi^2}{4\sqrt2}\right)^2\right]\ . 
 $$ 
A comparison of the width of $\mathcal{Y}$ with that of the distribution function, $\Delta\varepsilon \sim T$, 
leads to the criterion (\ref{cr}). 
The second order term in (\ref{S-phi-low-temp}) as well as the higher ones are small as $(T/E_{\rm Th})^{2n}$. 
Once we approach the singularity, i.e., once we reach the distance $\sqrt{\Phi^2 + \phi^2} \sim \sqrt{T/E_{\rm Th}}$,  then all the terms in the $\varphi_\varepsilon$ expansion of $\mathcal{Y}$ are of the same order and, as a result, all the terms in the expansion (\ref{S-phi-low-temp}) are of the order $\sim G_0 E_{\rm Th}$. The strong dependence 
on the direction from which the singularity is approached, i.e., on the 
angle $\theta\equiv \arcsin \frac{\phi}{\sqrt{\phi^2+\Phi^2}}$, in particular the vanishing of the leading term for $\theta=0$, is washed out in the higher order terms.
	
Close to the singularity point, i.e., for $\phi^2+\Phi^2 \ll  \frac{T}{E_{\rm Th}}$ the kernel $\mathcal{Y}$ as a function of 
$\varepsilon$ is more singular than the distribution function. This leads to our second main result: 
\begin{equation}
	S_{T\ll E_{\rm Th}} = 2\pi G_0 E_{\rm Th},  \quad  \phi^2+\Phi^2\ll   T/E_{\rm Th} \ .
	  	   \label{S-phi-low-temp-excess}
\end{equation} 
This expression follows from the $n=0$ delta-function in (\ref{y-1}) with the contributions of $n\neq 0$ being exponentially suppressed. As mentioned above, the strong dependence on the direction from which the singularity is approached, i.e., on the 
angle $\theta\equiv \arcsin \frac{\phi}{\sqrt{\phi^2+\Phi^2}}$, which is so prominent in the leading term of (\ref{S-phi-low-temp}), is washed out 
completely as we come close enough to the singularity, i.e., for $\Phi^2 + \phi^2 \ll T/E_{\rm Th}$.
	The excess noise (\ref{S-phi-low-temp-excess}) does not vanish at $T=0$. This may seem to be in conflict with the expected behavior of the equilibrium noise. The resolution of this apparent paradox is the fact that the area in the $\Phi,\phi$ plane, where this value of the noise is obtained shrinks to zero with $T\rightarrow 0$. 
		
Our main result is a strong enhancement of the noise near the singularity points $\Phi, \phi = 0\: ({\rm mod}\: 2\pi) $  in the quantum limit of $T\ll E_{\rm Th}$.
We use the term ``excess" here because the noise exceeds the value of the equilibrium Johnson-Nyquist noise in a single-channel normal conductor, $S_{\rm JN}=4G_0 T$. 	

In order to shed light on the origin of peculiar properties of the system (manifesting themselves in strong enhancement of the noise) at the singular points, we analyze the scattering states in this limit.
Let us consider the first and the last lines of the  left-contact scattering matrix, Eq.~ (\ref{s-matrix}), and of its right-contact counterpart. Absorbing $\alpha_i$ into the Aharonov-Bohm phase, we get the following  relations  for the  left and right Dirac-Majorana interfaces:
\begin{multline}
\begin{bmatrix}
	\psi_a(k) \\
	\psi_a^+(-k)
	\end{bmatrix}=\\=Q(\Phi,\phi,\varphi_k)\begin{bmatrix}
	\psi_b(k) \\
	\psi_b^+(-k)
\end{bmatrix}+q(\Phi,\phi,\varphi_k)\chi_l(k) ,
	\end{multline}
\begin{multline}
\begin{bmatrix}
	\psi_b(k) \\
	\psi_b^+(-k)
\end{bmatrix}=\\=Q(-\Phi,\phi,\varphi_k)\begin{bmatrix}
	\psi_a(k) \\
	\psi_a^+(-k)
\end{bmatrix}+q(-\Phi,\phi,\varphi_k)\chi_r(k),
	\end{multline}
with 
\begin{equation}
 Q(\Phi,\phi,\varphi_k)=\frac{e^{{\rm i}\varphi_k/2}}{2}\begin{bmatrix}
e^{{\rm i}\phi/2} & e^{{\rm i}\Phi/2}		  \\
	e^{-{\rm i}\Phi/2} & e^{-{\rm i}\phi/2}	
	\end{bmatrix},
\end{equation}
\begin{equation}
q(\Phi,\phi,\varphi_k) =\frac{{\rm i}e^{{\rm i }\varphi_k/4}}{\sqrt2}\begin{bmatrix}
	e^{{\rm i}\phi/4} \\
	-e^{-{\rm i}\phi/4}
\end{bmatrix}.
\end{equation}
Excluding $\psi_b$ and $\psi_b^+$ from these relations yields
\begin{multline}
\left[\mathbf{1}-Q(\Phi,\phi,\varphi_k)Q(-\Phi,\phi,\varphi_k)\right]	\begin{bmatrix}
		\psi_a(k) \\
		\psi_a^+(-k)
	\end{bmatrix}=\\=Q(\Phi,\phi,\varphi_k)q(-\Phi,\phi,\varphi_k)\chi_r(k)+q(\Phi,\phi,\varphi_k)\chi_l(k). \label{eq:1-qq}
\end{multline}
The determinant of the matrix in the square brackets is 
\begin{multline}
{\rm det}[\mathbf{1}-Q(\Phi,\phi,\varphi_k)Q(-\Phi,\phi,\varphi_k) ]=\\
=1-\frac{1}{2}e^{{\rm i}\varphi_k}(\cos\Phi+\cos\phi).  \label{eq:det}
\end{multline}
If the determinant is non-zero, the  fermion modes are the linear combinations of the incident $\chi_r$ and $\chi_l$, as described by the Eq.~(\ref{eq:Rk}). For the case of  zero determinant (\ref{eq:det}), which holds for $\varphi_k=2\pi n$, $\Phi=2\pi m$ and $\phi=2\pi l$,   eigenvalues of $[1-Q^2(2\pi n,2\pi m,2\pi l) ]$ are 1 and 0, with the corresponding eigenvectors being $\lambda_1=[1/\sqrt2; \ -1/\sqrt2]$ and $\lambda_0=[1/\sqrt2; \ 1/\sqrt2]$. These two vectors define Majorana modes $\eta$ and $\zeta$, which are eigenmodes  of the junction at the degeneracy points.  The vectors $\lambda_0$  and $\lambda_1$ correspond to the modes 
\begin{equation}\eta_{a,b}(k)\equiv \frac{1}{\sqrt2}[\psi_{a,b}(k)+ \psi_{a,b}^+(-k)]
\label{eta}
\end{equation}
 and 
\begin{equation}
\zeta_{a,b}(k)\equiv \frac{{\rm i}}{\sqrt2}[\psi_{a,b}^+(-k) - \psi_{a,b}(k)],
\label{zeta}
\end{equation}
respectively. 
Now we reformulate Eq. (\ref{eq:1-qq}) for the upper wire $a$ in the new basis of $\eta$ and $\zeta$:
\begin{multline}
\begin{bmatrix}
	\frac{1-e^{{\rm i }\varphi_k }}{\sqrt2} && \frac{{\rm i}}{\sqrt2}\\ \\
\frac{1-e^{{\rm i }\varphi_k }}{\sqrt2}	&& -\frac{{\rm i}}{\sqrt2}
\end{bmatrix}	\begin{bmatrix}
		\eta_a(k) \\ \\
		\zeta_a(k)
	\end{bmatrix}=\chi_l\frac{{\rm i}e^{{\rm i }\varphi_k/4}}{\sqrt2}\begin{bmatrix}
1 \\ \\
	-1
\end{bmatrix}. \label{eq:eta-zeta}
\end{multline}
For $\varphi_k \neq  2\pi l$  we obtain the solution of (\ref{eq:eta-zeta}) in the form
\begin{equation}
	\begin{bmatrix}
		\eta_a(k) \\ \\
		\zeta_a(k)
	\end{bmatrix}= \begin{bmatrix}
		0 \\ \\
		e^{{\rm i }\varphi_k/4}\chi_l(k)
	\end{bmatrix}. \label{eq:eta-zeta-1}
\end{equation}
At first glance, it may seem contradictory that the mode $\eta$ is absent for these values of $k$, i.e., a part of degrees of freedom is absent. 
What actually happens is a 
 redistribution of the continuous spectral weight of $\eta$   into the singular points of $\varphi_k = 2\pi l$.  Oppositely, the mode $\zeta_a$ has a constant spectral weight and does coincide with $\chi_l$ (up to the dynamical phase) and flows out in the right Majorana edge channel without a backscattering. The same holds for wire $b$ where $\zeta_b(k)=e^{{\rm i }\varphi_k/4}\chi_r(k)$.
Note, that from (\ref{eq:eta-zeta-1}) one obtains for the Dirac field  for a generic value of $\varphi_k \neq  2\pi l$
\begin{eqnarray}
\psi_a(k) &=& \frac{1}{\sqrt2}(\eta(k)+{\rm i}\zeta(k))=\frac{{\rm i}e^{{\rm i }\varphi_k/4}}{\sqrt2}\chi_l(k); \label{deg:psi} \\
\psi_a^+(-k) &=& \frac{1}{\sqrt2}(\eta(k)-{\rm i}\zeta(k))= - \frac{{\rm i}e^{{\rm i }\varphi_k/4}}{\sqrt2}\chi_l(k), \label{deg:psi-dagger}
\end{eqnarray} 
and thus $\psi_a(k)= - \psi_a^+(-k)$. Similarly, 
\begin{equation}\psi_b(k)= - \psi_b^+(-k)=\frac{{\rm i}e^{{\rm i }\varphi_k/4}}{\sqrt2}\chi_r(k).\label{deg:psi-b}
\end{equation}
Hence, particles and holes in a particular    Dirac channel  are not independent but rather form Majorana particles.   This correlation results in  zero values of the cumulants of the form  
\begin{multline}\langle\!\langle\psi_a^+(k)\psi_a(k)\psi_a^+(p)\psi_a(p)\rangle\!\rangle=\\=\frac{1}{2}\delta_{k,p}\Big[\langle\chi_l(-k)\chi_l(k)\rangle \langle\chi_l(k)\chi_l(-k)\rangle -\\
		- \langle\chi_l(k)\chi_l(-k)\rangle \langle\chi_l(-k)\chi_l(k)\rangle \Big]=0.
	\end{multline} 
 At the same time, the fermions in different wires are fully independent, $\langle\psi_a^+(k)\psi_b(k)\rangle=0$, as follows from Eqs.~ (\ref{deg:psi}) and (\ref{deg:psi-b}). Consequently, there is no contribution to the noise from the off-resonant states.

 Let us discuss the redistribution of the spectral weight of the new modes $\eta$ and $\zeta$. To this end, we first calculate the spectral weight for arbitrary values of the phases and then consider a transition to the singular limit, $\Phi\to 0$ and $\phi\to 0$. Using the $2\times 2$-matrix $\mathbf{R}_k$, Eq.~(\ref{eq:Rk}), and definitions  (\ref{eta}) and (\ref{zeta}), we obtain for a given $k$:
\begin{equation}
	\begin{bmatrix}
		\eta_a(k)\\
		\zeta_a(k)\\
		\eta_b(k)\\
		\zeta_b(k)
	\end{bmatrix}=
	\begin{bmatrix}
		x_{1,k}(\Phi,\phi)  && x_{2,k}(\Phi,\phi) \\
		z_{1,k}(\Phi,\phi)  && z_{2,k}(\Phi,\phi) \\
		x_{2,k}(-\Phi,\phi)  && x_{1,k}(-\Phi,\phi) \\
		z_{2,k}(-\Phi,\phi)  && z_{1,k}(-\Phi,\phi) 
	\end{bmatrix}\begin{bmatrix}
		\chi_l(k)\\ 
		\chi_r(k) 
	\end{bmatrix}. \label{eq:eta}
\end{equation}
The matrix elements in the above formula are related to $r_{1,2}$ from Eq.(\ref{eq:r1r2}) as
\begin{multline}
	x_{j,k}(\Phi,\phi)=\frac{1}{\sqrt2}[r_j(\Phi,\phi,k)+r_j^*(\Phi,\phi,-k)], \label{eq:uv} \\
	z_{j,k}(\Phi,\phi)=\frac{-{\rm i}}{\sqrt2}[r_j(\Phi,\phi,k)-r_j^*(\Phi,\phi,-k)],  \ {j=1,2}.
\end{multline}
The   scattering amplitudes $x$ and $z$  determine the density of states for the new modes in $a$ and $b$ wires: 
\begin{equation}
	\rho_{\eta,a}(\Phi, \phi,\varepsilon)= |x_{1,\varepsilon}(\Phi,\phi)|^2 + |x_{2,\varepsilon}(\Phi,\phi)|^2,
\end{equation}
\begin{equation}
	\rho_{\zeta,a}(\Phi,\phi,\varepsilon)= |z_{1,\varepsilon}(\Phi,\phi)|^2 + |z_{2,\varepsilon}(\Phi,\phi)|^2,
\end{equation}
and   $$
\rho_{\eta,b}(\Phi,\phi,\varepsilon)=\rho_{\eta,a}(-\Phi,\phi,\varepsilon), 
$$
$$ \rho_{\zeta,b}(\Phi,\phi,\varepsilon)=\rho_{\zeta,a}(-\Phi,\phi,\varepsilon).
$$
Calculating the limit of $\Phi\to 0$ and $\phi\to 0$, keeping $\varphi_\varepsilon$ fixed and non-zero, one obtains that the mode $\zeta $ has constant density of states,
$$
\rho_{\zeta,a}(\varepsilon)=1,
$$
while the mode $\eta$ has a singular spectral weight located at the Andreev levels,
$$
\rho_{\eta,a}(\varepsilon) = \sum\limits_n \delta(\varphi_\varepsilon-2\pi n).
$$
 It should be emphasized that both singular amplitudes are equal to each other,  $$|z_{1,\varepsilon}(0,0)|^2=|z_{2,\varepsilon}(0,0)|^2=\frac{1}{2}\sum\limits_n \delta(\varphi_\varepsilon-2\pi n).$$ This means that the resonant mode $\eta$, which propagates in both of the Dirac wires, is an equal-amplitude superposition of $\chi_l$ and $\chi_r$ at the discrete energies $\varepsilon_n$. Such a resonant  correlation between the  Dirac states in $a$ and $b$ wires---which should be contrasted to the case of $\varepsilon \neq  \varepsilon_n$---is responsible for the noise enhancement.

\subsection{Equilibrium noise at the degeneracy point}
\label{Results:Equilibrium_noise_Arbitrary_temperature}

In this Section we generalize our result for the noise at the degeneracy point $\Phi=\phi=0$ for arbitrary temperatures.
From (\ref{noise-0}) and (\ref{y-1}) we obtain
\begin{equation}
	S_{\rm deg}(T)= 4\pi G_0 E_{\rm Th } \sum\limits_{n=-\infty}^\infty \frac{1}{\cosh  \frac{2\pi n E_{\rm Th}}{T} +1}\ ,   \label{P-1}
\end{equation}
where $S_{\rm deg} \equiv S(\Phi=0,\phi=0)$.
As already mentioned, in the low temperature limit the  only term with $n=0$ survives and we obtain  (\ref{S-phi-low-temp-excess}). In the opposite limit of high temperature, $T\gg E_{\rm Th}$, we replace the summation by an integral over ${\rm d}x=2\pi E_{\rm Th}/T$  and obtain the thermal noise  $S_{\rm JN} $ like in a normal  channel:
\begin{equation}
\lim_{\frac{T}{E_{\rm Th}}\to \infty}	T^{-1}S_{\rm deg}(T)= 4  G_0\ .   \label{P-2}
\end{equation}

\subsection{Equilibrium noise: High temperature regime }
\label{Results:Equilibrium_noise_High_temperature}
Below we obtain the $\phi$-dependent finite temperature correction to $S_{\rm JN}$ at the regime   $T\gg E_{\rm Th} $  and zero Josephson current ($\Phi=0$). 
For this case the zero-frequency noise is given by 
\begin{eqnarray}
	S_{\Phi=0}(\phi,T)
	= G_0   \int  \frac{{\rm d}\varepsilon}{\cosh  \frac{\varepsilon }{T} +1} Y(\phi,\varphi_\varepsilon),  \label{c2-1} \\
	Y(\phi, \varepsilon)=\frac{    (1-\cos \phi)(1+\cos \frac{\varepsilon}{E_{\rm Th}}) }{ 1 + \frac{(1+\cos  \phi)^2 }{4}   -(1+\cos  \phi)\cos  \frac{\varepsilon}{E_{\rm Th}}}    \ ,
\end{eqnarray}
  where
$
Y(\phi,\varphi_\varepsilon)=\mathcal{Y}(\Phi=0,\phi,\varphi_\varepsilon).
$ 
At high temperatures the distribution function decays smoothly with the energy while $Y(\phi,\varphi_\varepsilon)$ rapidly oscillates. We, thus, expand the kernel into a cosine series $$Y(\phi,\varepsilon)=\sum\limits_0^\infty y_n(\phi) \cos \frac{n\varepsilon}{E_{\rm Th}}$$  and retain only the zeroth and the first Fourier harmonics 
\begin{equation}Y(\phi,\varepsilon)\approx y_0(\phi)+y_1(\phi)\cos\frac{\varepsilon}{E_{\rm Th}}. \label{y-2}\end{equation}
We obtain 
 $
y_0(\phi)=2
$ 
and
$ 
y_1(\phi)=3+\cos \phi.
$ 
The zeroth harmonic yields the usual Johnson-Nyquist noise as leading term while the first harmonic gives a small $\phi$-dependent correction: 
\begin{multline}
	S_{\Phi=0}(T{\gg} E_{\rm Th},\phi)=4 G_0 T+  \frac{2\pi (3+\cos \phi)  G_0T^2}{ E_{\rm Th}\sinh\frac{\pi T}{E_{\rm Th}}}    .	\label{S-high-temp}
\end{multline}
This  correction is exponentially suppressed for $T\gg E_{\rm Th} $ similar to the critical current in (\ref{CPhR}). In this case the thermal length is shorter than the normal region perimeter and the superconducting correlations and interference of Andreev pairs are destroyed by thermal fluctuations.

\section{Conclusions}
\label{Discussion}

In this work we have studied the equilibrium zero-frequency noise in a chiral link between two topological superconductors.
This system can be realized on a surface of a 3D topological insulator covered with superconducting and magnetic films. 
Our system is a single-channel ballistic Josephson junction where the Andreev pairs can be thought of as 
being the scattering states of the incident chiral Majorana fermions in the leads. We have derived the effective action for the full counting statistics of the charge transfer in the Majorana representation.  We have shown that for temperatures lower than the Thouless energy $k_{\rm B}T\ll E_{\rm Th}$ the system is characterized in equilibrium by an excess zero-frequency noise with the maximum of $S=  2 \pi G_0 E_{\rm Th}$, which is large compared to the thermal noise in a normal channel  $ S_{\rm JN}=4G_0 k_{\rm B}T $. Here $G_0=e^2/(2\pi \hbar)$. Moreover, we have obtained oscillations of the noise power as a function of the superconducting phase bias and  the Aharonov-Bohm flux. The dependence on the AB flux has a fractional $h/e$ period because of the chiral nature of the split conducting channels.  
The large noise is a consequence of  the emergent ground state degeneracy at {\it even}  Aharonov-Bohm and superconducting phases  of $2\pi n $.  The current-phase relation is also singular at these points. This is distinct from non-topological SNS contacts, where spikes are possible for {\it odd} phases  of $\pi(2n+1)$. Hence, the singularities at even phases can be considered as a signature of the gapless Majorana leads. 

 It is instructive to compare our result with that obtained for ordinary ballistic SNS junctions \cite{PhysRevLett.76.3814,PhysRevB.53.R8891}.
There, a large noise was obtained as a result of rare switching events (telegraph noise) between two Andreev levels \cite{PhysRevLett.76.3814}.
  For low temperatures, this occurs in a tiny region around   $\pi(2n+1)$  phases; otherwise the noise is  exponentially suppressed. In contrast,  in our interferometer the noise is never exponentially suppressed: away from the degeneracy it saturates to $S_{\rm JN}$.  The enhancement of noise around the degeneracy of Andreev levels in  the ordinary ballistic SNS junctions and in our Majorana interferometer suggest a certain similarity between the two mechanisms.
   It should be emphasized, however, that the analysis of Refs.~\cite{PhysRevLett.76.3814,PhysRevB.53.R8891} requires introduction of a rate $\delta$ of inelastic transitions between the Andreev levels induced by coupling to an external bath (in practice, phonons). In our problem, a counterpart of this $\delta$ is the width of the spectral-weight peak given by Eq.~(\ref{w1}).

An important conclusion about the real part of the low frequency impedance  of the  junction considered in this work follows from the fluctuation-dissipation theorem, $S = 4T {\rm Re} [Z^{-1}]$. We observe that the inverse impedance of the junction must have a 
large real part at the low temperatures in addition to the usual inductive (imaginary) part describing the Josephson effect. 
Namely
\begin{equation}
\frac{1}{Z(\omega\rightarrow 0)}=\frac{1}{Z_J(\omega)} + \frac{1}{Z_{\rm diss}}\ ,
\end{equation}
where $Z_J = - i\omega L_J$ and $L_J = (2\pi/\Phi_0)(\partial I(\Phi,\phi)/\partial \Phi)$ is the Josephson inductance.
Our result means for the dissipative part that the following estimates hold at $T \ll E_{\rm Th}$
\begin{eqnarray}
&&\frac{1}{Z_{\rm diss}} \sim G_0\quad {\rm for }\quad \Phi^2 + \phi^2 \gg T/E_{\rm Th}\ , \nonumber\\
&&\frac{1}{Z_{\rm diss}} \sim G_0 \frac{E_{\rm Th}}{T}\quad {\rm for }\quad \Phi^2 + \phi^2 \ll T/E_{\rm Th}\ .
\end{eqnarray}
Thus, our Josephson contact can be thought of as a parallel connection of a Josephson element and a resistive shunt, 
whose conductance is strongly dependent on the phases $\Phi$ and $\phi$.

\section{   Acknowledgments } This research was financially supported by the DFG-RSF grant (No. 16-42-01035 (Russian node) and No. SH 81/4-1, MI 658/9-1  (German node)).

\end{document}